\documentclass[12pt]{iopart}

\usepackage{graphicx}

\def\raa{$R_{\rm AA}$}
\def\rs{\sqrt{s_{\rm NN}}}
\begin{document}

\title{Particle Production at Large
Transverse Momentum with ALICE}

\author{Harald Appelsh\"auser$^1$ for the ALICE Collaboration}

\address{$^1$Institut f\"ur Kernphysik, Goethe-Universit\"at, Max-von-Laue-Str. 1,
60438 Frankfurt am Main, Germany}
\ead{appels@ikf.uni-frankfurt.de}
\begin{abstract}
We present transverse momentum distributions of inclusive charged particles and identified hadrons in $pp$ and Pb--Pb collisions at $\rs= 2.76$~TeV, measured by ALICE at the LHC. The Pb--Pb data are presented in intervals of collision centrality and cover transverse momenta up to 50 GeV/$c$. Nuclear medium effects are studied in terms of the nuclear modification factor \raa. The results indicate a strong suppression of high-$p_T$ particles in Pb--Pb collisions, consistent with a large energy loss of hard-scattered partons in the hot, dense and long-lived medium created at the LHC. We compare
the preliminary results for inclusive charged particles to previous results from RHIC and
calculations from energy loss models. Furthermore, we compare the nuclear modification
factors of inclusive charged particles to those of identified $\pi^0$, $\pi^{\pm}$, K$^0_s$,
and $\Lambda$.
\end{abstract}


\section{Introduction}
The first discussion of jet quenching in Quark-Gluon Plasma dates back almost 30 years. In a paper by Bjorken~\cite{bjorken}, elastic scattering 
from "quanta in the plasma" was considered as a possible mechanism for differential
energy loss of high-energy quarks and gluons, leading to a modification of the jet structure
in high-energy hadronic collisions. Ten years later first quantitative predictions were given
in a perturbative QCD-based approach~\cite{wang1}, indicating that the dominant energy loss mechanism is
radiative and leads to a significant suppression of hadron production in central nucleus-nucleus collisions at high transverse momentum $p_T$.
Another decade later, at Quark Matter 2001, the first experimental evidence for jet quenching
was presented by the RHIC collaborations~\cite{qm01}.
It is remarkable that these very early data already revealed the essence of the phenomenon,
which is valid to date: At large transverse momentum, hadrons are suppressed by a factor five
in central Au--Au collisions at top RHIC energy. Over the past ten years, very detailed experimental
information has been collected at RHIC, such as the suppression pattern of many different
particle species including photons, the reaction plane dependence, and the important $pp$ 
and d-Au reference data. All these results substantiate the early findings.
On the theoretical side the situation is less clear. There exist quite a few models which
give a good description of the experimental results, however, the extraction of the relevant
medium properties from these data is still ambiguous (see e.g.\,\cite{majumder-vLeeuwen} for
a recent review).

At this Quark Matter conference, yet another ten years after the experimental discovery 
of jet quenching at RHIC, first results from LHC are presented. Owing to the increase
of collision energy by more than an order of magnitude as compared to RHIC, the production
of high-$p_T$ particles at LHC is much more abundant, giving rise to the expectation
that LHC data on particle production at high $p_T$ will substantially improve the experimental
constraints on energy loss models and allow for a more precise determination of the medium 
parameters. A first impression was given by ALICE presenting the inclusive
charged particle spectra
in central Pb--Pb collisions at $\rs = 2.76$~TeV out to 
$p_T=20$~GeV/$c$, exhausting the present $p_T$
reach at RHIC after only a few days of minimum bias data taking at the LHC~\cite{alice-raa}.

In this contribution, preliminary results on particle production at high $p_T$, 
measured with ALICE at the LHC are presented. This includes a centrality dependent
study of inclusive charged particles in Pb--Pb collisions at $\rs =2.76$~TeV out to 
$p_T=50$~GeV/$c$, 
and studies of nuclear modification of charged and neutral pion, K$^0_s$, and $\Lambda$
production, employing different particle identification techniques in ALICE~\cite{kalweit-qm11}.
The results are based on data from the first heavy-ion
campaign at the LHC in the fall of 2010, where ALICE recorded about 30\,M minimum bias Pb--Pb events.

Nuclear modifications of particle production in AA are commonly expressed in terms of the nuclear
modification factor \raa:
\begin{equation*}
R_{\rm AA}=
\frac{{\rm d}N^{AA}/{\rm d}p_T}{\langle N_{\rm coll}\rangle {\rm d}N^{\rm pp}/{\rm d}p_T}.
\end{equation*}
The construction of $R_{\rm AA}$ requires a reference $p_T$ distribution in $pp$ which
is obtained from the analysis of a dedicated $pp$ run at $\sqrt{s}=2.76$~TeV at
the LHC in March 2011, where ALICE collected about 70\,M minimum bias events.
The $pp$ reference is scaled to the proper centrality in Pb--Pb by the mean number
of nucleon-nucleon collisions $\langle N_{\rm coll}\rangle$ which is obtained from
a Monte-Carlo Glauber model implementation~\cite{toia}.

\section{Results}
The results presented here are mainly based on the analysis of charged particle
tracks reconstructed in the central tracking detectors ITS (Inner Tracking System) and
TPC (Time Projection Chamber). The combined 
momentum resolution of the TPC-ITS tracking system was verified with high momentum
cosmic muon tracks and invariant mass distributions of K$^0_s$ and $\Lambda$. 
The $p_T$ resolution can be parametrized by
$\sigma_{p_T}/p_T \approx 0.002 \cdot p_T$, yielding about 10\%
at $p_T=50$~GeV/$c$. More details about the ALICE detector can be found in~\cite{alice-jinst}.

\begin{figure}[tbp] 
   \centering
  \includegraphics[width=3.in]{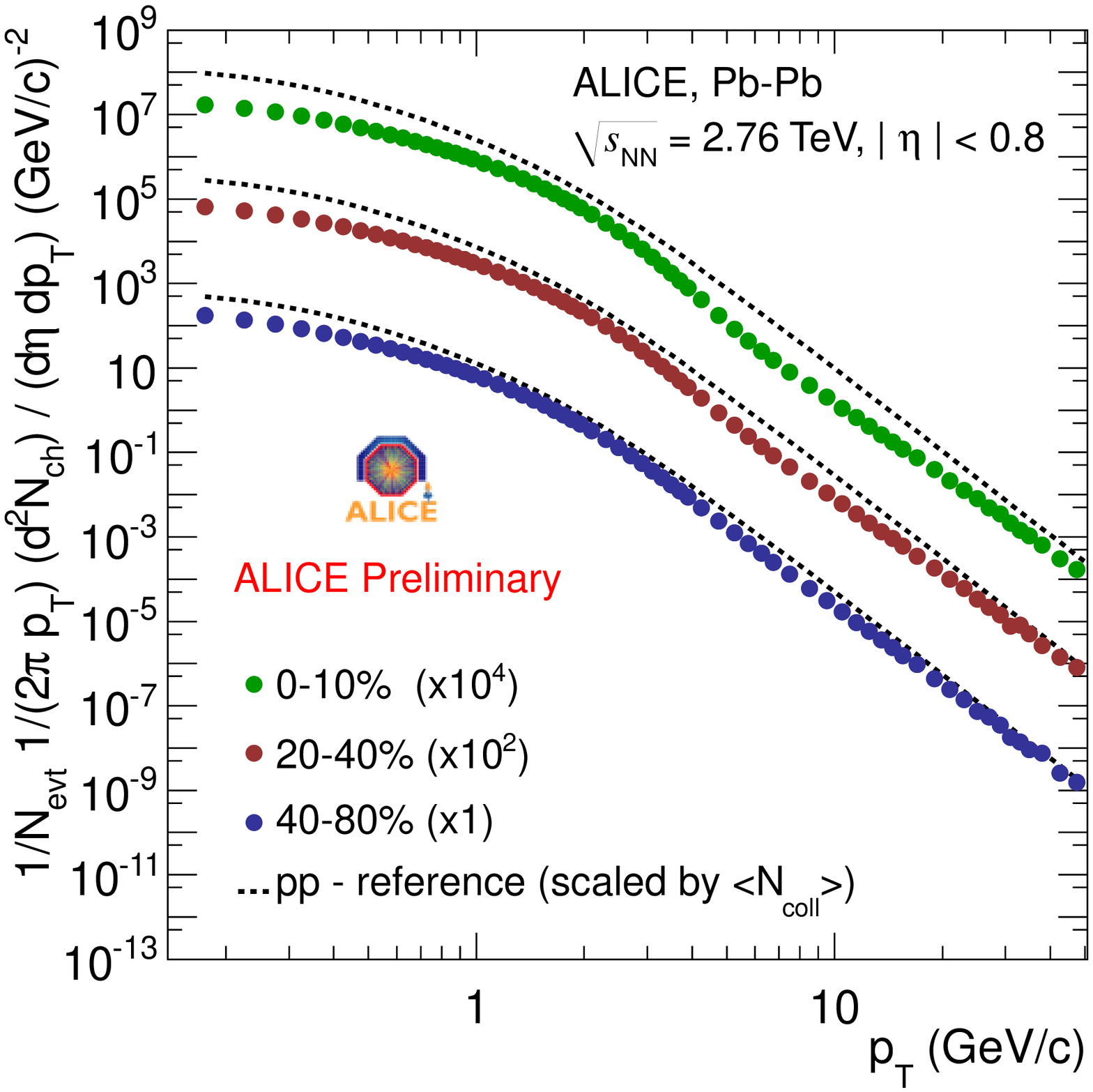}   
      \includegraphics[width=3.in]{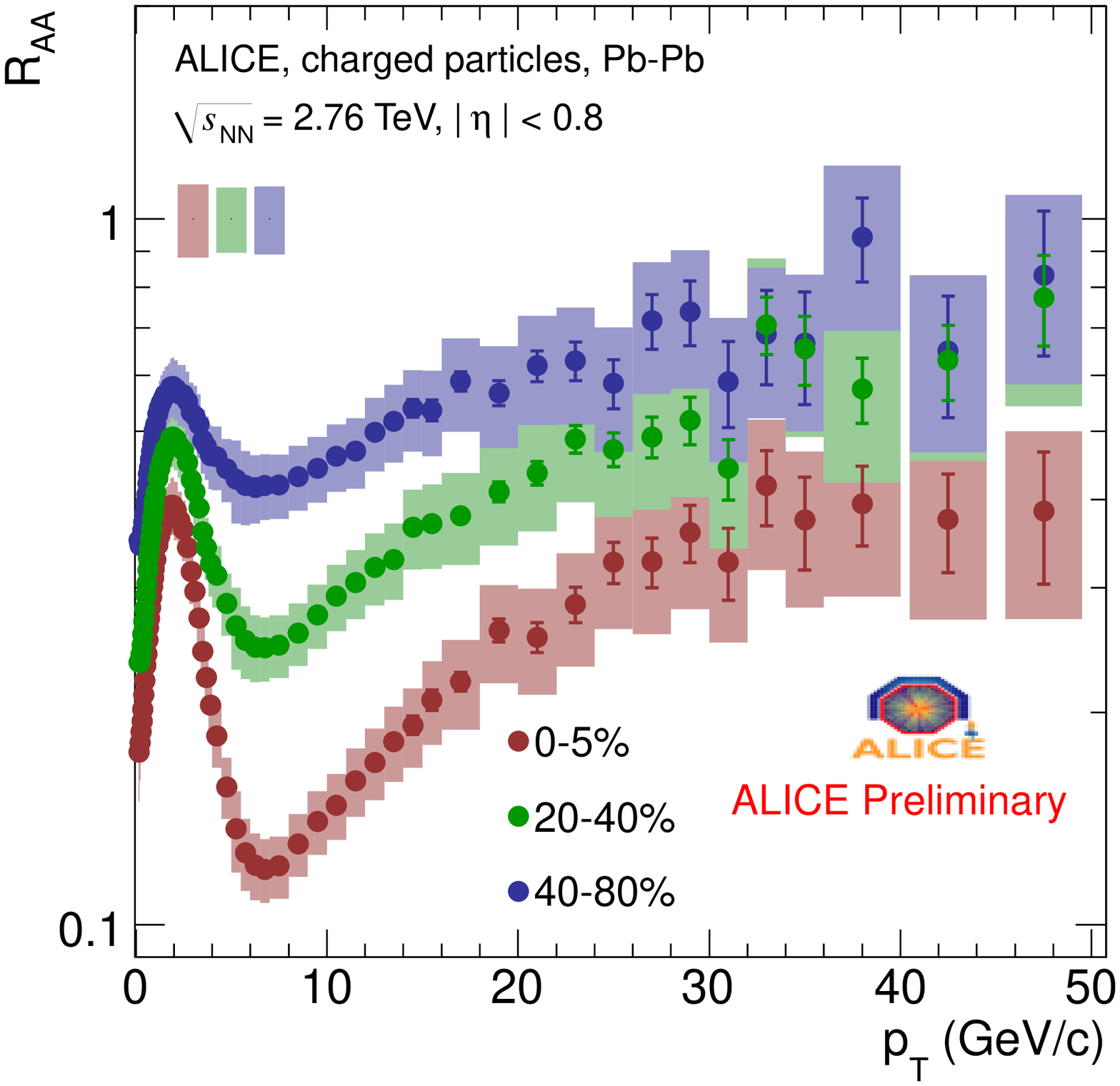} 
   \caption{Left: Differential invariant yields of inclusive
   charged particles in Pb--Pb collisions at $\rs = 2.76$~TeV.
   Results from different centrality intervals are separated by successive factors of $10^2$
    for clarity. Also shown (dashed lines) are the reference spectra derived from $pp$ collisions at 
    $\rs=2.76$~TeV and scaled by $ \langle N_{\rm coll}\rangle$. Right: \raa\,of inclusive charged particles. Error bars indicate the statistical uncertainties. The error boxes contain the
    systematic uncertainties on the Pb--Pb data and on the $pp$ reference. Normalization uncertainties
     are indicated by the bars at \raa$=1$.}
   \label{pb-spec}
\end{figure}

\subsection{Inclusive charged particles}
Figure~\ref{pb-spec} (left panel) shows the corrected $p_T$ distributions of 
inclusive charged particles
in Pb--Pb collisions at $\rs = 2.76$~TeV in different
centrality intervals~\cite{jacek-qm11}. 
Also shown are the $pp$ reference spectra derived from the $pp$
data at $\sqrt{s}=2.76$~TeV~\cite{jacek-qm11,knichel-qm11} and scaled by 
$ \langle N_{\rm coll}\rangle$~\footnote{After the conference presentation, we found small inconsistencies in the tracking, leading to an overestimation of \raa\,at high transverse momenta. The analysis was reviewed, and here we present the updated result in the transverse momentum range $p_{T}< 50$~ GeV/$c$, where the updated values remain within the systematic uncertainties of the ones shown at the conference.}.
Due to limited statistics, the measured yield in $pp$ 
is used only below $p_T=6$~GeV/$c$ for the reference. At larger $p_T$ the $pp$ reference 
was approximated by a modified Hagedorn function~\cite{hagedorn}. 
This functional form provides the best fit to
the measured $pp$ spectrum in $6<p_T<40$~GeV/$c$ and was extrapolated to $p_T=50$~GeV/$c$. The estimated uncertainty on the $pp$ reference due to the parametrization
and extrapolation procedure increases with $p_T$ and reaches 20\% at $p_T=50$~GeV/$c$.
The scaled $pp$ reference spectra reveal the typical
power law shape at high $p_T$, characteristic for hard scattering. In contrast, 
a marked 
depletion of the Pb--Pb spectra is developing gradually as centrality is increasing, indicating
a significant suppression of particle production in central Pb--Pb collisions.

The nuclear modification factors \raa\,out to $p_T=50$~GeV/$c$ are shown in Fig.~\ref{pb-spec}
(right panel)
for different centrality intervals. At all centralities, a pronounced minimum at about $p_T=6$--8~GeV/$c$
is observed, above which \raa\,rises monotonically up to about $p_T=50$~GeV/$c$.

The nuclear modification factor \raa\,in central Pb--Pb collisions (0--5\%) obtained with the measured
$pp$ reference 
is slightly (10--20\%) below the results previously 
published in~\cite{alice-raa} where
the $pp$ reference was based on an interpolation procedure using $pp$ data at
$\sqrt{s}=0.9$ and 
7~TeV measured by ALICE.  The new results are within the uncertainties quoted in~\cite{alice-raa}. The new reference based on measured $pp$ data at $\sqrt{s} = 2.76$~TeV leads to a significant improvement of the systematic uncertainties at $p_T<20$~GeV/$c$. This substantiates the previous findings in~\cite{alice-raa} that the suppression observed in central collisions at the LHC is stronger than at RHIC~\cite{star,phenix}.

\begin{figure}[tbp] 
   \centering
   \includegraphics[width=3in]{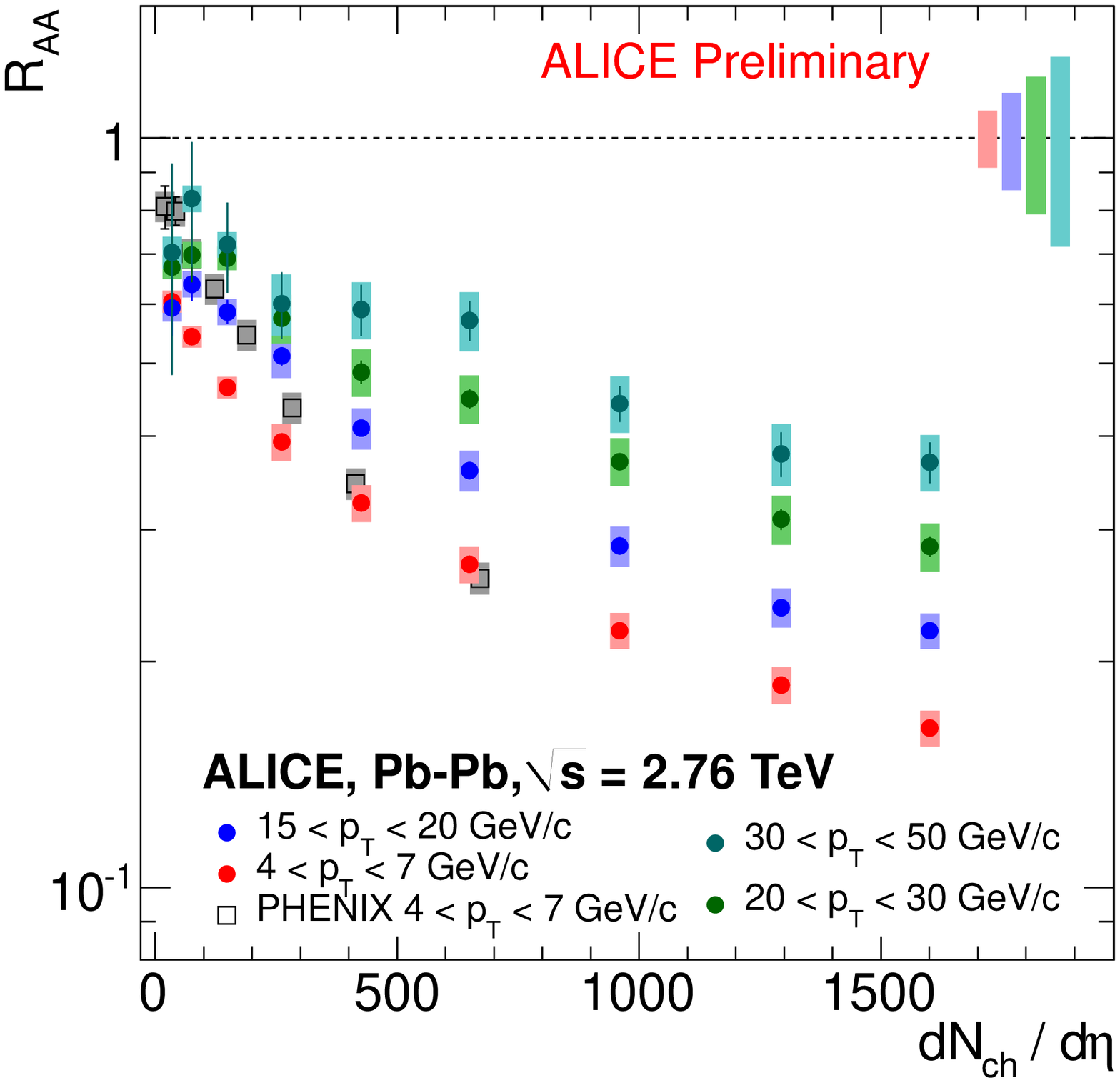} 
   \includegraphics[width=3in]{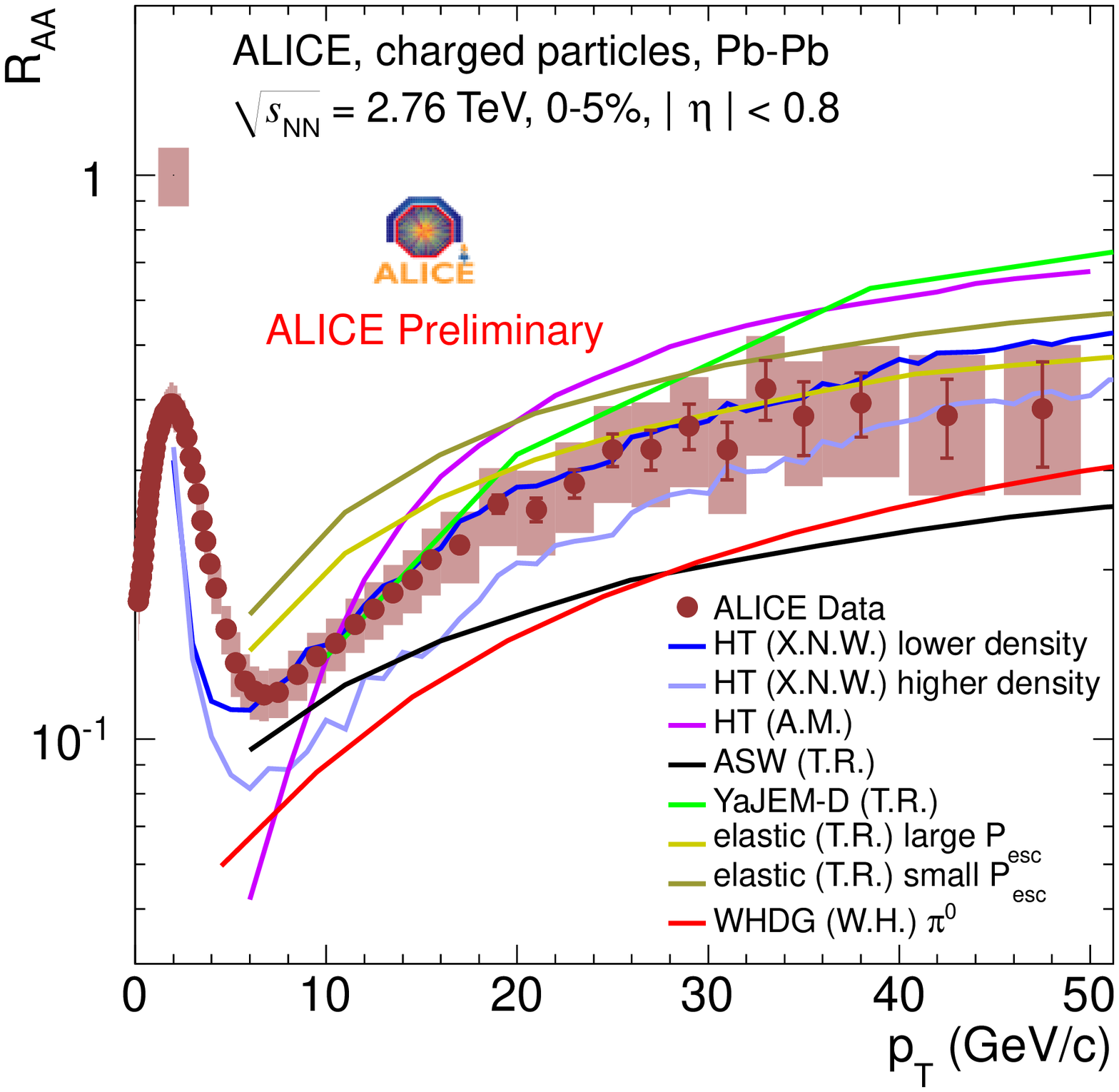} 
\caption{Left: \raa\,in four $p_T$ intervals as a function of the charged particle density
${\rm d}N_{\rm ch}/{\rm d}\eta$. Error bars indicate the statistical uncertainties. The boxes
contain the systematic errors on the Pb--Pb data and on $ \langle N_{\rm coll}\rangle$. The centrality-independent uncertainties on the $pp$ reference are indicated by the boxes
at \raa$=1$. Also shown are results from PHENIX at RHIC~\cite{phenix}. Right: \raa\,in 
central (0--5\%) Pb--Pb collisions compared to model 
calculations~\cite{xnw,majum,renk,horo}.}
   \label{raa-cent-dndeta}
\end{figure}

The centrality dependence of \raa\,is illustrated in Fig.~\ref{raa-cent-dndeta} (left panel)
where the yields have been integrated over ranges in $p_T$ and the resulting 
nuclear modification factors are plotted as a function of the charged particle density
${\rm d}N_{\rm ch}/{\rm d}\eta$ (from~\cite{alice-multi-cent}). In the range
$4<p_T<7$~GeV/$c$ we observe the strongest centrality dependence. Also shown
are results from the PHENIX experiment at lower collision energy which show a compatible
dependence on ${\rm d}N_{\rm ch}/{\rm d}\eta$. This suggests that charged particle density
is indeed a key parameter for the jet quenching mechanism, as implied by most of the energy loss models. Deviations at very small particle density may be related to collision geometry and
will have to be studied further.
As $p_T$ increases, the centrality dependence of \raa\,decreases gradually.

Figure~\ref{raa-cent-dndeta} (right panel) shows a comparison of \raa\,in central (0--5\%) Pb--Pb collisions
at $\rs =2.76$~TeV to calculations from energy loss models~\cite{xnw,majum,renk,horo}. 
All model calculations
 in Fig.~\ref{raa-cent-dndeta} have been constrained to  match \raa\,results from RHIC.
The qualitative features
of our data are described by all models, including the strong rise of \raa\,for $p_T>7$~GeV/$c$.
A quantitative comparison of the model calculations
to the present data will help to put tighter constraints on the underlying energy loss
mechanisms and their parametric dependencies. The less steeply falling $p_T$ spectrum
of hard-scattered partons at the LHC as compared to lower collision energies increases
the sensitivity to the details of the parton energy loss distribution $P(\Delta E)$, whereas a steep 
spectrum is mainly sensitive to the escape probability $P(\Delta E=0)$~\cite{renk}.

\begin{figure}[tbp] 
   \centering
   \includegraphics[width=3in]{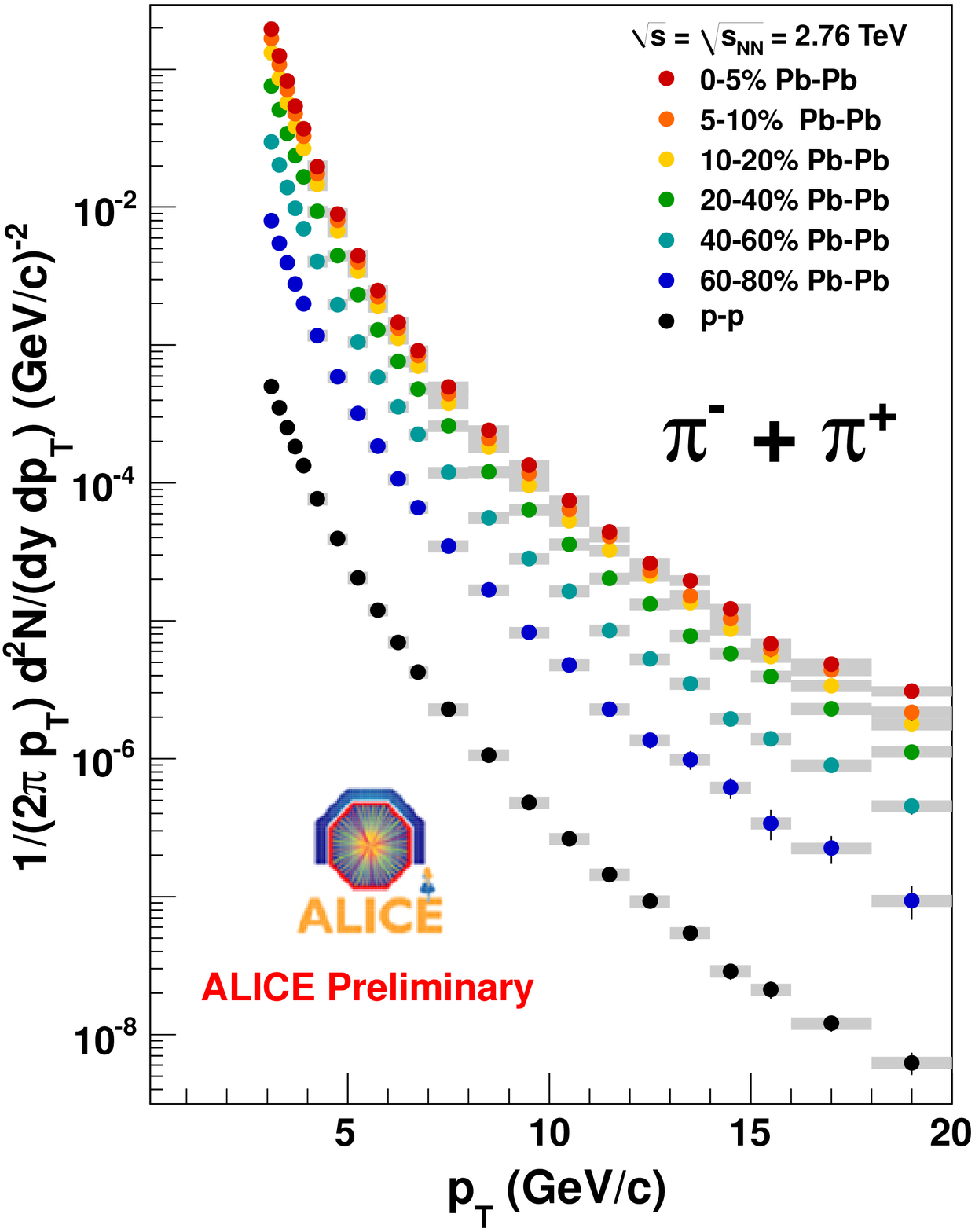} 
   \includegraphics[width=3in]{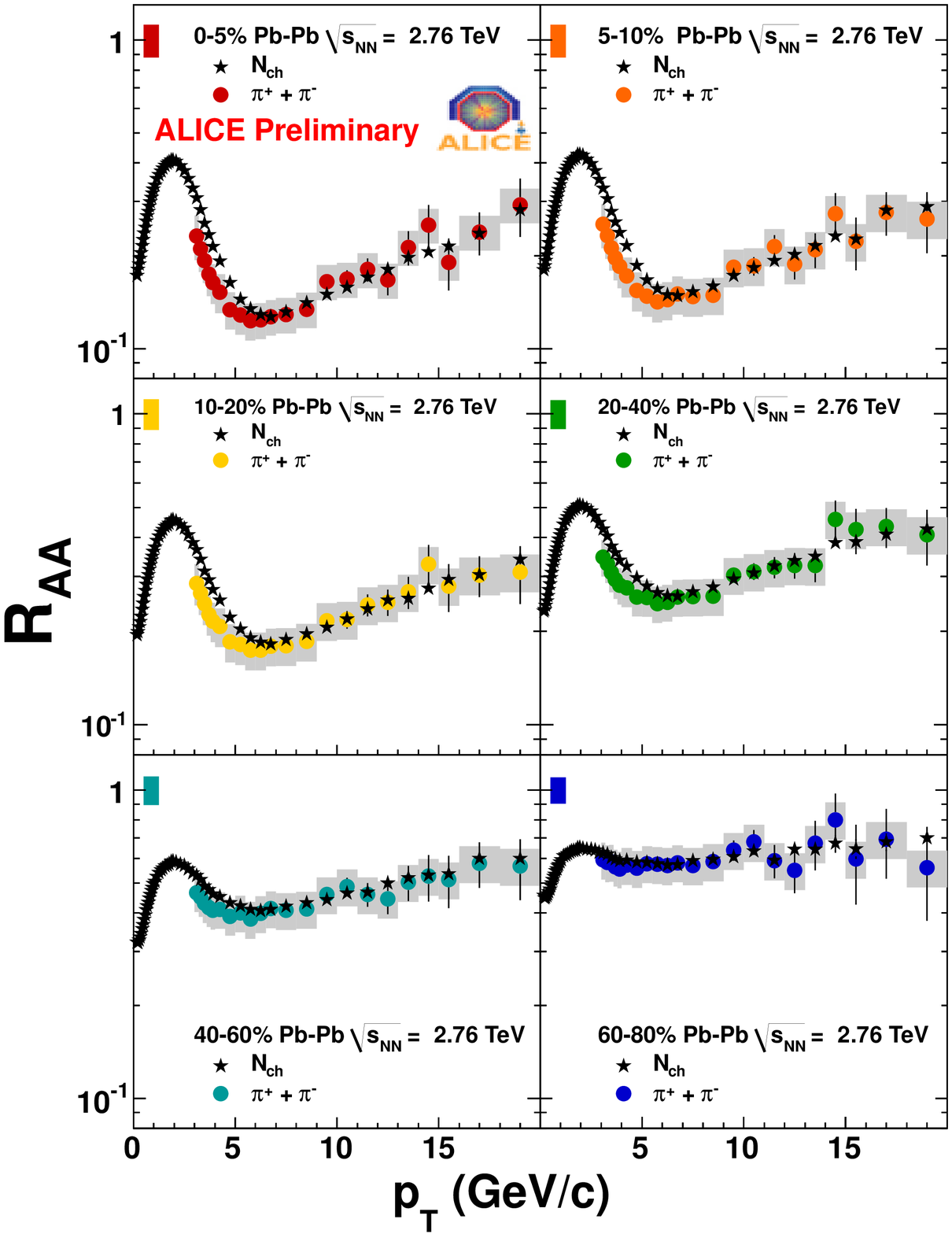} 
  \caption{Left: Differential invariant $p_T$ yield of charged pions in $pp$ and Pb--Pb collisions at
  $\rs =2.76$~TeV. Right: \raa\,of charged pions compared to inclusive charged particles. Errors are
    as defined in Fig.~\ref{pb-spec} (right).}
   \label{spectra-picharged}
\end{figure}

\subsection{Charged pions}

The analysis of charged pions at high $p_T$ is based on statistical particle
identification using the specific energy loss d$E$/d$x$ in the TPC~\cite{christiansen-qm11}. 
In the region of the relativistic rise of the energy loss ($p_T>3$~GeV/$c$) the separation
of pions from kaons and protons is nearly independent of $p_T$ out to $p_T=50$~GeV/$c$.
The fraction of pions from all charged particles is determined in bins of $p_T$ by
fitting the d$E$/d$x$ distribution with four Gaussians for p, K, $\pi$, and e. Small 
corrections arise from the contribution of muons and the slightly different reconstruction
efficiencies for pions and charged particles. For the measurement of \raa, the analysis was performed in minimum bias $pp$ collisions and in bins of centrality in Pb--Pb
collisions, both at $\sqrt{s_{\rm NN}}=2.76$~TeV. Figure~\ref{spectra-picharged} (left panel)
shows the differential invariant $p_T$ yields of charged pions at mid-rapidity 
($|\eta|<0.8$) in $pp$ and Pb--Pb collisions at $\rs=2.76$~TeV. 
The charged pion \raa\,is shown in Fig.~\ref{spectra-picharged} (right panel) and 
compared to the \raa\,of inclusive charged particles. At $p_T<6$~GeV/$c$, the charged pion \raa\,is slightly
below that of the charged particles. This effect increases with increasing centrality and 
may be related to enhanced baryon production in AA collisions, as was observed 
at RHIC.  At high $p_T$ 
($p_T>6$~GeV/$c$), the charged pion \raa\,is compatible with the \raa\,of all the charged particles.

\begin{figure}[tbp] 
   \centering
   \includegraphics[width=3in]{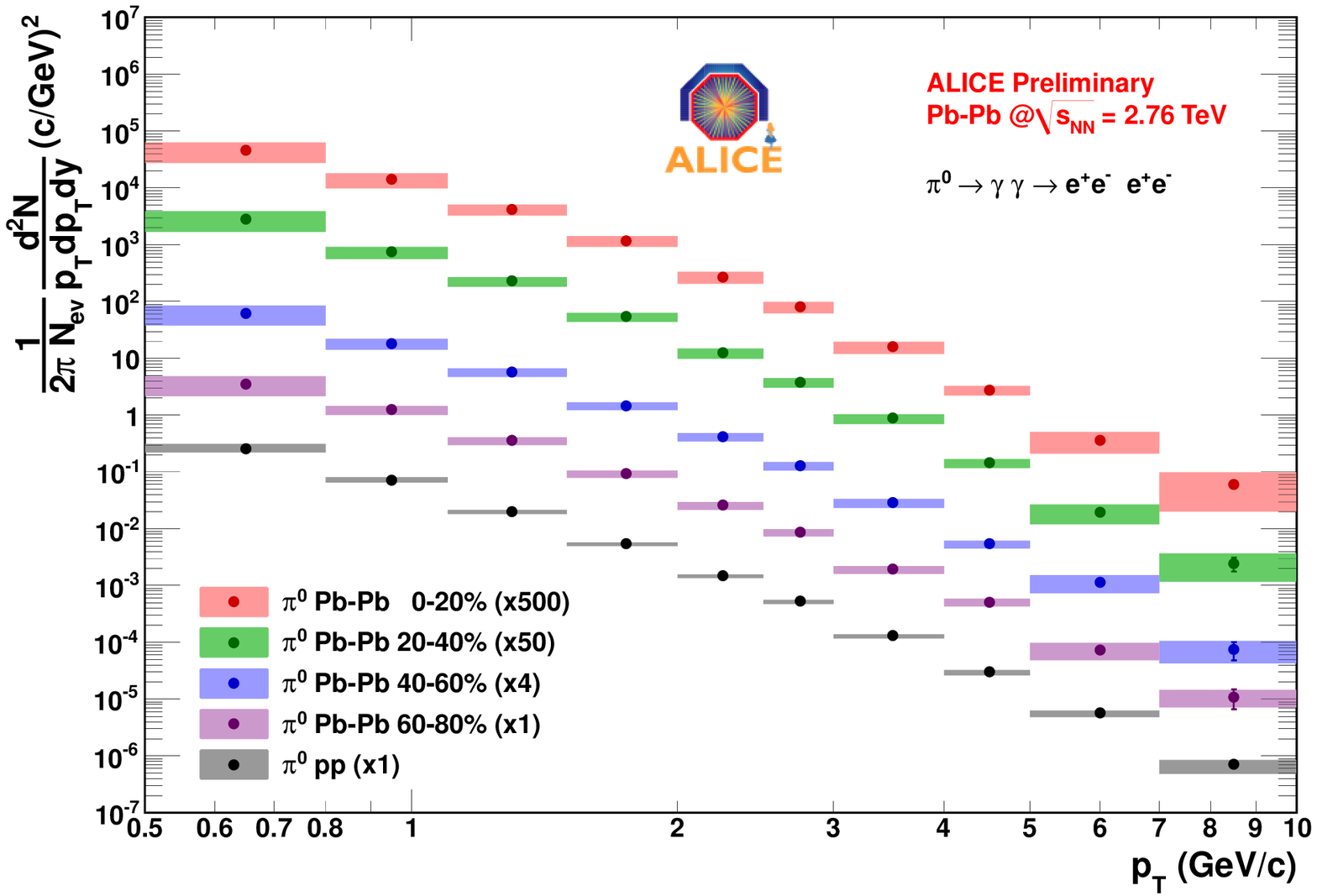} 
   \includegraphics[width=3in]{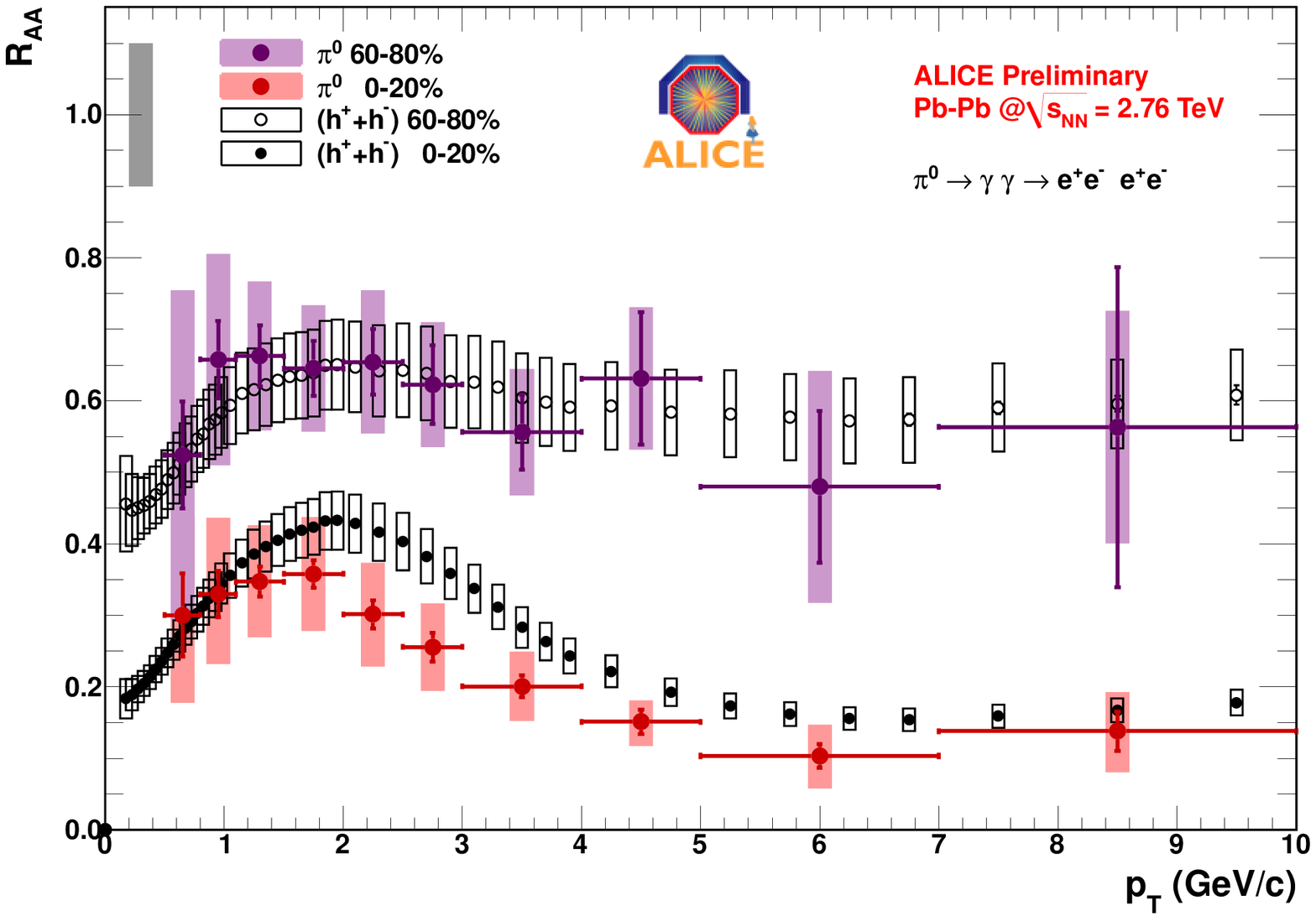} 
   \caption{Left: Differential invariant yield of $\pi^0$ in $pp$ and Pb--Pb collisions at
  $\rs=2.76$~TeV obtained from $\gamma$ conversions. 
  Right: \raa\,of $\pi^0$ compared to inclusive charged particles. Errors are as defined  
    in Fig.~\ref{pb-spec} (right).}
   \label{spec-raa-pizero}
\end{figure}

\subsection{Neutral pions and strange particles}
The topological reconstruction of secondary decay vertices of neutral particles 
allows the identification of $\pi^0\rightarrow\gamma\gamma$ via conversion
of the decay photons~\cite{reygers-qm11,gustavo-qm11}. The conversion method is complementary to calorimetric
measurements and extends the acceptance for $\pi^0$ to low $p_T$. 
The invariant yields of $\pi^0$ in $pp$ and Pb--Pb collisions at $\rs =2.76$~TeV 
and the nuclear modification factors \raa\,are shown in Fig.~\ref{spec-raa-pizero}.
The \raa\,of $\pi^0$ is compatible with that of inclusive 
charged particles in peripheral Pb--Pb collisions. In central events, the \raa\,of $\pi^0$ is slightly below that of charged particles, most pronounced in the range
$2<p_T<5$~GeV/$c$. This is consistent with the results for charged pions
described above.

\begin{figure}[tbp] 
   \centering
   \includegraphics[width=3.in]{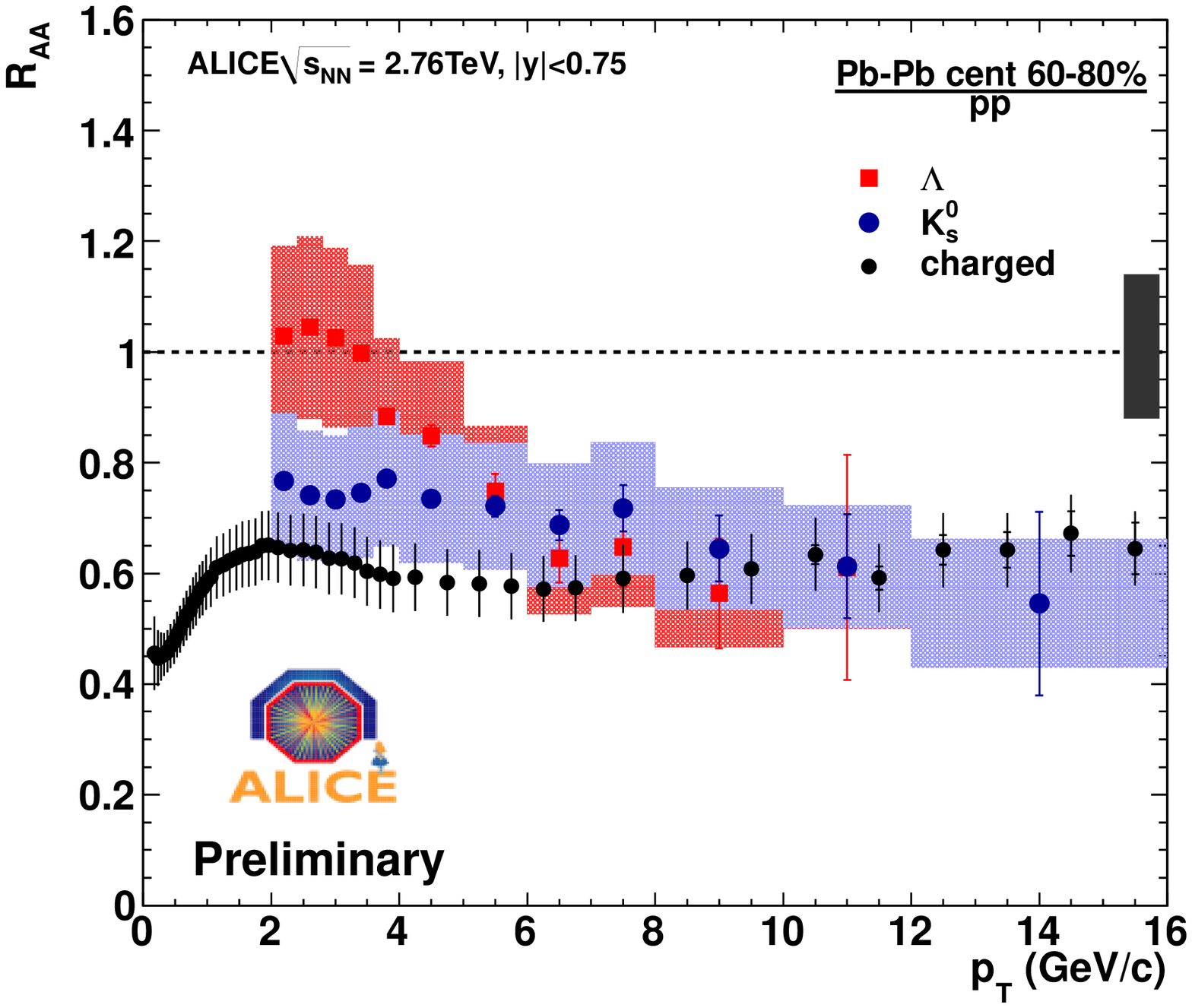} 
   \includegraphics[width=3in]{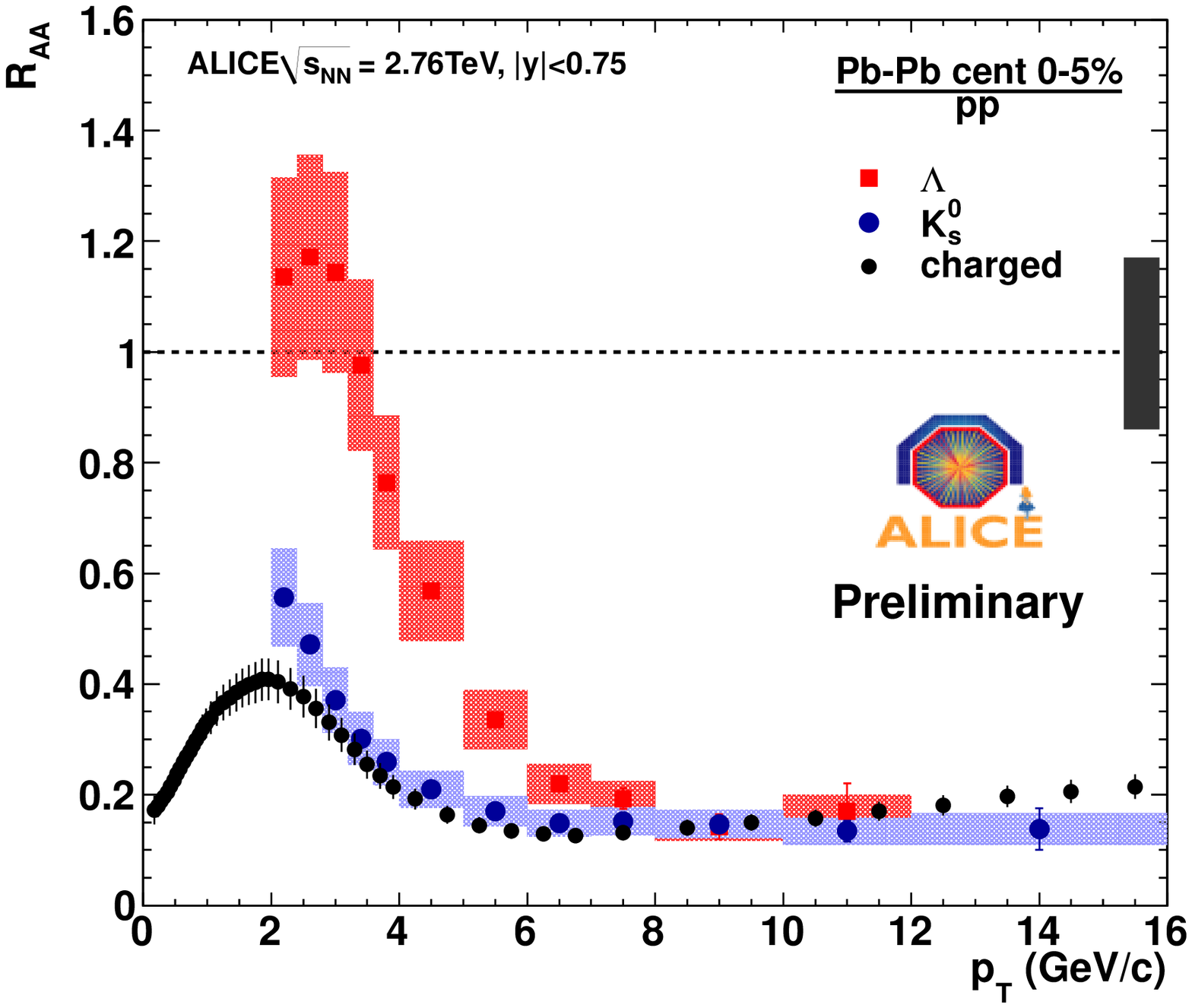} 
   \includegraphics[width=3.in]{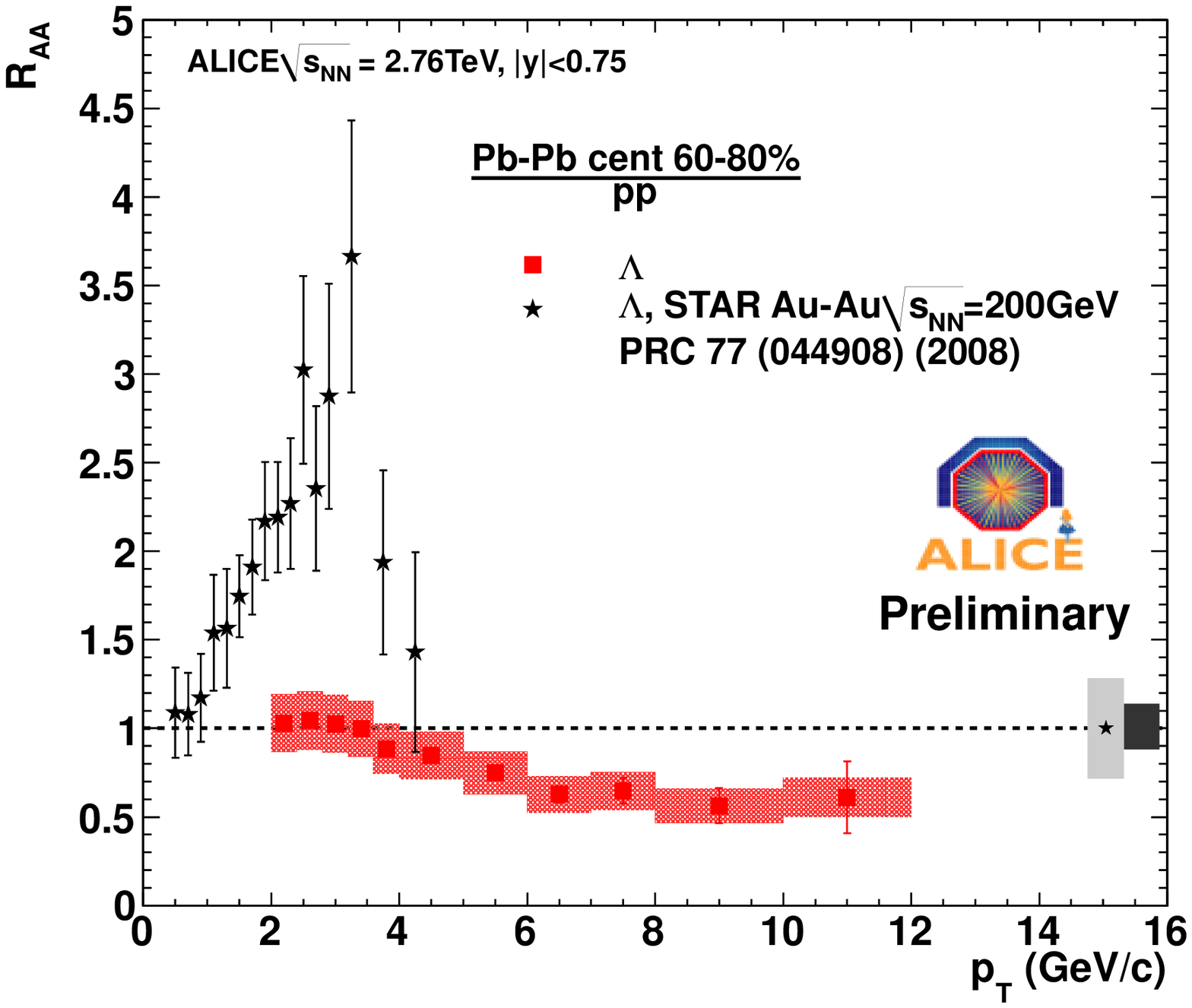} 
   \includegraphics[width=3in]{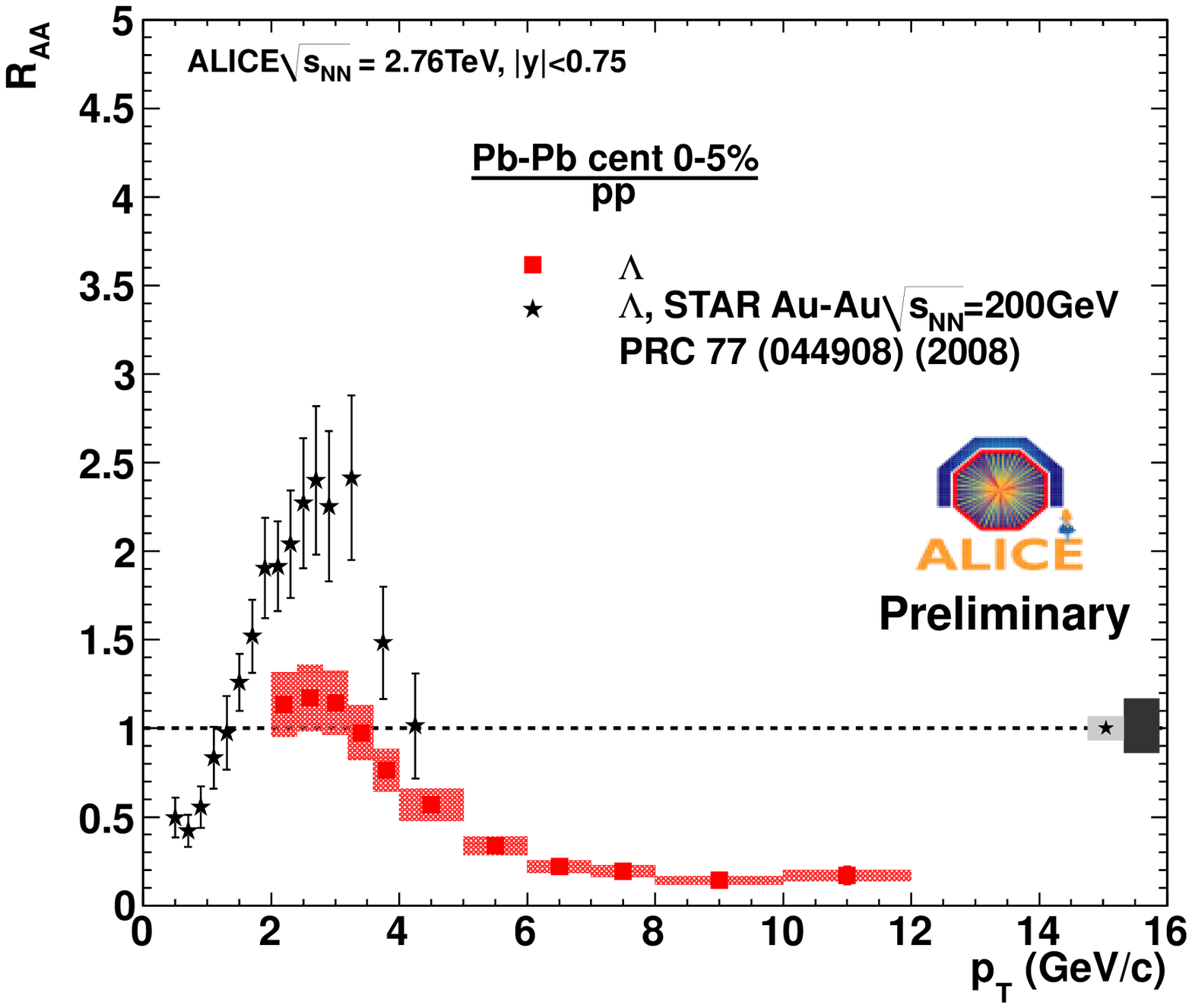} 
   \caption{Upper row: \raa\,of $\Lambda$ and K$^0_s$ in peripheral (left) and central (right) 
   Pb--Pb collisions at 2.76~TeV. The results are compared to inclusive charged particles.
    Errors are as defined in Fig.~\ref{pb-spec} (right). Lower row: \raa\,of $\Lambda$ compared
    to results from STAR at RHIC~\cite{star-lambda}.}
   \label{strange-raa}
\end{figure}

The reconstruction of weak decays  $\Lambda\rightarrow{\rm p}+\pi^-$ and 
K$^0_s\rightarrow\pi^++\pi^-$ at high $p_T$~\cite{simone-qm11} allows to 
study different suppression
patterns for baryons and mesons, which may give a handle on how to separate quark
and gluon energy losses. On the upper panels of Fig.~\ref{strange-raa}, \raa\,for 
K$^0_s$ and $\Lambda$ are shown in comparison to inclusive charged particles
out to $p_T=16$~GeV/$c$.
In peripheral events, the \raa\,of K$^0_s$ shows very little $p_T$ dependence and
is compatible with \raa\,of charged particles. The \raa\,of
$\Lambda$ is significantly above that of charged particles at $p_T<6$~GeV/$c$, reaching
unity at $2<p_T<4$~GeV/$c$. This might be related with enhanced baryon production
in AA collisions. At $p_T>6$~GeV/$c$ both K$^0_s$ and $\Lambda$ are compatible
with charged particles.
In central collisions the observations are qualitatively similar but more dramatic. 
For $p_T>6$~GeV/$c$, a significant suppression for K$^0_s$ and $\Lambda$ is seen which is similar to that of inclusive charged particles.
At lower $p_T$, the \raa\,of $\Lambda$ is significantly larger than
that of K$^0_s$ which is in line with the observation of a strong and centrality dependent
enhancement of $\Lambda$/K$^0_s$~\cite{iouri-qm11}.

The lower panels of Fig.~\ref{strange-raa} show a comparison of the \raa\,of $\Lambda$ to results from STAR~\cite{star-lambda} at lower collision energy, where a significant enhancement above unity was reported in peripheral and central collisions. Such an enhancement is not seen in ALICE. 
We note that the difference between the collision energies is similar in peripheral and
central collisions, implying that $R_{\rm CP}$ is compatible~\cite{simone-qm11}. This points
to a change of $\Lambda$ production in $pp$ between RHIC and LHC.

\section{Summary}
Measurements of particle production at high $p_T$ in Pb--Pb collisions at $\rs=2.76$~TeV are presented for the first time at a Quark Matter conference. The new data from LHC 
provide unprecedented experimental information on jet quenching in heavy-ion collisions. 
The results presented by ALICE indicate strong suppression of charged particles in 
Pb--Pb collisions and a characteristic centrality and $p_T$ dependence of the nuclear
modification factor \raa. The suppression observed in central Pb--Pb collisions
at $\sqrt{s_{\rm NN}} = 2.76$~TeV at the LHC
is stronger than in central Au--Au collisions at $\sqrt{s_{\rm NN}} = 0.2$~TeV at RHIC. 
However, we find that the results are 
compatible if compared in terms of the charged particle density ${\rm d}N_{\rm ch}/{\rm d}\eta$.
This emphasizes the strong relation between medium density and partonic energy loss.
A comparison to model calculations indicates a high sensitivity of high-$p_T$ LHC data
to details of the energy loss mechanism. This gives rise to the expectation that
these data will provide stringent constraints to the energy loss models and pave the way for a detailed extraction of the relevant medium parameters. 
Additional information on partonic energy loss is provided by the measurement of
nuclear modification factors of identified hadrons. Charged and neutral pions give
consistent results and indicate a high-$p_T$ suppression similar to that of inclusive charged
particles. 
The study of strange particles $\Lambda$ and K$^0_s$ produced in heavy-ion collisions
was extended to a $p_T$ range beyond the reach of RHIC, out to $p_T=16$~GeV/$c$.
We observe that the different particle species show individual suppression patterns which 
may be related to enhanced baryon production in heavy-ion collisions.
At $p_T>8$~GeV/$c$, the suppression of all hadron species under study becomes universal.

\end{document}